\documentclass[12pt]{iopart}
\usepackage{graphicx}
\usepackage{color}

\usepackage{hyperref}
\hypersetup{
    colorlinks=true, 
    linktoc=all,     
    linkcolor=blue,  
}
\usepackage{placeins}
\usepackage{booktabs}
\usepackage{blindtext}
\makeatletter
\newcommand{\colorcaption}[2][]{%
  \begingroup%
  \renewcommand{\@caption@fignum@sep}{ (Color online). }%
  \caption[#1]{#2}%
  \endgroup%
}
\makeatother
\begin{document}

\title{Minimal Theory of Isomerism- Q.Q and Other Interactions}
\author{}
\author{P. C. Srivastava$^a$   and  L. Zamick$^{b}$}
\address{$^a$Department of Physics, Indian Institute of Technology, Roorkee 247 667, 
India}
\address{$^b$Department of Physics and Astronomy, 
 Rutgers University, Piscataway, New Jersey 08854, USA }

\date{\today}
\begin{abstract}
We perform shell model calculations using a quadrupole-quadrupole
interaction (Q.Q) in a single  $j$ shell space. We show that this one-parameter
interaction is a good predictor of where nuclear isomerism occurs
and where it does not occur. The limitations of this interaction are
also discussed. We then include other interactions, e.g. in the f$_{7/2}$
shell those obtained from the (local) spectrum of a two particle system
and then from a two hole system. In the g$_{9/2}$ where there is insufficient
empirical data from the two hole system and so various empirical interactions
are used. 
\end{abstract}


\section{Introduction}

{\color{black} Study of a nuclear isomer in the different region of nuclear chart attracted much experimental and theoretical investigations \cite{axel,pcbh,jain1,Maheshwari1,jain2}.}
In the present work we consider two types of isomeric
states. In one case, we have the spin-gap isomer state of angular
momentum J for which there are no states of angular momentum (J-1)
or (J-2) below the {\color{black} isomeric state}. These isomers can have very long half-lives.
 They cannot decay by E1, M2 or E2 transitions. Then there
is the second kind {\color{black}of isomer}, where there is a, say, a state of angular momentum
(J-2) below the J state but the energy difference is so small that
the transition is hindered; e.g. for E2 {\color{black} transitions the rate goes as
($E_{i}$ -$E_{f}$)$^{5}$; for $M_1$ and $E_1$ as ($E_{i}$-$E_{f}$)$^{3}$.}

{\color{black}While admirable} progress has been made in the calculations of larger and
larger spaces for quantitative shell model properties {\color{black} \cite{pcs2,pcs,Bharti}}, it should not
be forgotten that much insight can be gained by doing simple calculations
with simple interactions. We here discuss, in part, the quadrupole-quadrupole
interaction (Q.Q) which has only one parameter i.e. the overall strength.
This parameter can be adjusted to give the correct excitation energy.
We will show that the Q.Q interaction is a good predictor
of where isomeric states can be found.
It should be said that the idea of using matrix elements from experiment
came from earlier works of deShalit and Talmi \cite{Shalit,Talmi}.

We then consider, in addition, other empirical (fitted) interactions. In the
f$_{7/2}$ shell we can use the spectrum of the two-particle system
$^{42}$Sc {\color{black}from the experimental data for the effective interactions} to make predictions of isomerism in $^{43}$Sc, $^{43}$Ti and $^{44}$Ti. We then use, the local hole-hole spectrum of $^{54}$Co
to make predictions of the isomerism in $^{53}$Fe, $^{53}$Co and
$^{52}$Fe. In the g$_{9/2}$ shell, there is insufficient {\color{black}experimental} data about
the two hole system $^{98}$In so empirical two-body matrix elements are
obtained by other means \cite{Coraggio,Qi,Flowers}.

For a neutron and proton in a single j shell, the states of even angular
momentum J have isospin T= 1 whilst the odd ones have T=0 zero. The
T=0 two-body matrix elements can only be obtained from the np system-
not from nn or pp. They are usually not as well known as the ones
with T=1. Because of charge independence the spectra of T=1 states
for the $np$, $nn$ and $pp$ systems are nearly identical and this is a big
help in obtaining the corresponding two-body matrix elements {\color{black} extracted from the energy-levels}.
We will later show the striking differences in T=0 matrix elements obtained
in 1964 \cite{Zamick16,Bayman,McCullen,Ginocchio} and the ones obtained in 2006 \cite{Escuderos} when better
data were available. {\color{black} Our aim is to see differences in the energies
with earlier and new set of effective interactions.}

Isomeric states played  a very prominent role in the ground breaking
book, elementary theory of nuclear shell structure,  by Mayer and
Jensen \cite{Mayer}. They are cited as evidence that there is indeed a
shell structure in nuclei. For updated reading and broadening the
scope the review articles by Heyde and Wood \cite{Heyde} on shape isomers
and by Draculis, Walker and F.G. Kondev \cite{Kondev} on isomers in heavier
nuclei are recommended. The latter also discusses medical applications
of nuclear isomers.

\noindent %
\noindent\begin{minipage}[t]{1\columnwidth}%
Table 1: {\color{black} Two-body matrix elements of the Q.Q interaction. {\color{black} We have chosen $\chi^\prime$ =1 corresponding to
$\chi^\prime$ = $\chi~b^4$.} }
\end{minipage}

\hspace{-1cm} %
\begin{tabular}{|c|c|c|c|c|c|c|c|c|c|c|}
\hline 
J  & 0  & 1  & 2  & 3  & 4  & 5  & 6  & 7  & 8  & 9\tabularnewline
\hline 
\hline 
f$_{7/2}$  & -1.9184  & -1.5530  & -0.8952  & -0.0914  & 0.6395  & 1.0049  & 0.6396  & 0.8952  &  & \tabularnewline
\hline 
g$_{9/2}$  & -2.9178  & -2.5642  & -1.9010  & -1.0168  & -0.0442  & 0.8400  & 1.4147  & 1.4147  & 0.5305  & -1.5916\tabularnewline
\hline 

\end{tabular}

We will be dealing mainly with high spin isomers. In the single j
shell it is easy to determine the maximum J. For identical particles
put a given particle in the highest m state consistent with the Pauli
principle. For example, for $^{43}$Sc the one proton is in  m=7/2 state and
the two neutrons in 7/2 and 5/2. This adds up to 19/2 which is indeed
J$_{max}$ for $^{43}$Sc. It is also J$_{max}$ for $^{43}$Ti,
$^{53}$Co and $^{53}$Fe. For $^{44}$Ti we have two protons and two neutrons one with
m=7/2 and one with m=5/2. {\color{black} This is responsible to generate $J=12^+$.}

\noindent %
\noindent\begin{minipage}[t]{1\columnwidth}%
Table 2: Two body matrix elements used in the f$_{7/2}$ shell.%
\end{minipage}

\hspace{-1cm} %
\begin{tabular}{|c|c|c|c|c|c|c|c|c|}
\hline 
J  & 0  & 1  & 2  & 3  & 4  & 5  & 6  & 7\tabularnewline
\hline 
\hline 
MBZ(1964)  & 0.0000  & 1.036  & 1.509  & 2.248  & 2.998  & 1.958  & 3.400  & 0.617\tabularnewline
\hline 
MBZE (2006)$^{42}$Sc  & 0.0000  & 0.6110  & 1.5803  & 1.4904  & 2.8153  & 1.5100  & 3.2420  & 0.6163\tabularnewline
\hline 
hole-hole $^{54}$Co  & 0.0000  & 0.9369  & 1.4457  & 1.8215  & 2.6450  & 1.8770  & 2.9000  & 0.1974\tabularnewline
\hline 
Q.Q  & 0.0000  & 0.3655  & 1.0232  & 1.8270  & 2.5579  & 2.9233  & 2.5580  & 1.0232\tabularnewline
\hline 
\end{tabular}

\vspace{10mm}

\noindent %
\noindent\begin{minipage}[t]{1\columnwidth}%
Table 3: Two-body matrix elements used in the g$_{9/2}$ shell. %
\end{minipage}

\hspace{-1.5cm} %
\begin{tabular}{|c|c|c|c|c|c|c|c|c|c|c|}
\hline 
J  & 0  & 1  & 2  & 3  & 4  & 5  & 6  & 7  & 8  & 9\tabularnewline
\hline 
\hline 
CCGI  & 0.000 & 0.829  & 1.710  & 1.877  & 2.217  & 2.046  & 2.383  & 1.913  & 2.527  & 0.915\tabularnewline
\hline 
Qi et al.  & 0.000  & 1.220  & 1.458  & 1.592  & 2.283  & 1.882  & 2.549  & 1.930  & 2.688  & 0.626\tabularnewline
\hline 
g$_{9/2}$ Q.Q  & 0.000  & 0.3536  & 1.0168  & 1.8990  & 2.8736  & 3.7618  & 4.3325  & 4.3325  & 3.4483  & 1.3262\tabularnewline
\hline
INTd  & 0.000  & 1.1387  & 1.3947  & 1.8230  & 2.0283  & 1.9215  & 2.2802  & 1.8797  & 2.4275  & 0.7500\tabularnewline
\hline 
\end{tabular}

\section{The Q.Q and other interactions in the single j shell}

 The interaction we use is -$\chi$ Q.Q = -$\chi$ $\sqrt{5}$ {[}(r$^{2}$
Y$^{2}$)$_{i}$(r$^{2}$ Y$^{2}$)$_{j}${]}$^{0}$.
Two body matrix elements were constructed using
harmonic oscillator radial wavefunctions.
These have one parameter, the oscillator length $b$, which is approximately
equal to $A^{1/6}$ $fm$. In evaluating energies, unless specified otherwise, we set $\chi$b$^{4}$ to 1 MeV. Alternately, one can say that
the energy is in units of $\chi$b$^{4}$.
 Results for two-body matrix elements of this interaction are shown in table 1.

We shall also be showing results from other interactions for comparison.
In the early 1960's empirical two body matrix elements were taken from the
spectrum of $^{42}$Sc in order to do calculations in the f$_{7/2}$ region.
We cite the works of Bayman et al. \cite{Bayman}, McCullen et al. \cite{McCullen}
and Ginocchio and French \cite{Ginocchio}. However, at that time (1964) the
T=0 two body matrix elements were not well determined. The T=1 states
also occurred in $^{42}$Ca and $^{42}$Ti and so were much better
known. 
{\color{black} Despite these deficiencies we show in the second column of table 2 the results of the old MBZ (1964) interaction. 
This should be compared with the results of the newer MBZE (2006) interaction shown in the 3rd column,
as performed by Escuderos, Zamick and Bayman \cite{Escuderos}. The latter calculation is basically the same as the 1964 one 
except that the input parameters were better known in 2006.}
Also, shown are matrix elements
from the two hole system $^{54}$Co. These are appropriate for nuclei
in the upper part of the f$_{7/2}$ shell (had one used the same interaction,
the spectrum of holes would be the same as that for particles). In
the first 3 rows of table 2 the ground state energy has been set to
zero. To make a better comparison with Q.Q we also added a constant
(1.9184 MeV) to the matrix elements in the first row of Table 1 and
show this in the last row of Table 2. Adding a constant will not affect
the level spacings.  {\color{black} In table 3 we have shown the  two-body matrix elements used for the 
$g_{9/2}$ shell calculations.}


We note that besides a strongly attractive J=0, T=1 matrix element,
Q.Q has also attractive matrix elements for the neutron-proton system
in the T=0 channel, namely for J=1 and J=J$_{max}$, the latter being
seven in the f$_{_{7/2}}$shell and nine in the g$_{9/2}$ shell. This is
also a feature of the empirical two-body matrix elements in both shells.
The Q.Q interaction is thus quite different from the J=0 T=1 pairing
interaction which was in vogue in the early fifties, e.g. in the works
of Flowers \cite{Flowers} and Edmund and Flowers \cite{Edmonds}.

\section{The spectra with a Q.Q and other interactions.}
In this section we present results of single j shell calculations
of energy levels for selected nuclei in both the f$_{7/2}$ and g$_{9/2}$
regions. These are contained in Table 4 to 9. In Table 4, we show the
spectra of even-even nuclei using the Q.Q interaction; in Table 5,
odd A nuclei are considered and in Table 6, odd-odd nuclei. In Tables
7,8 and 9 we have reported the corresponding spectra, but using local interactions
from experiment. For the lower part of the f$_{7/2}$ shell we use
the particle-particle spectrum of $^{42}$Sc as input whilst in the
upper half the hole-hole spectrum of $^{54}$Co. Discussions will
follow in the next section.  In tables 4 to 8 we give results for $\chi^{''}=$$\chi$b$^{4}$ = 1. 
 A reasonable fit {\color{black} with the experimental data} in the $f_{7/2}$ and $g_{9/2}$ region is given for $\chi$ =0.4 MeV. \\


\noindent\begin{minipage}[t]{1\columnwidth}
Table 4: Energy levels (in MeV)  of even-even nuclei with a Q.Q interaction. Energy are in MeV. %
\end{minipage}

 \hspace{2cm}
\begin{tabular}{|c|c|c|c|c|}
\hline 
J  & $^{44}$Ti ,$^{52}$Fe  & $^{48}$Cr  & $^{96}$Cd  & $^{92}$Pd,$^{88}$Ru\tabularnewline
\hline 
\hline 
0  & 0.000  & 0.000  & 0.000  & 0.000\tabularnewline
\hline 
1  & 2.995  & 2.296  & 5.040  & 4.747 \tabularnewline
\hline 
2  & 0.570  & 0.552  & 0.867  & 0.563 \tabularnewline
\hline 
3  & 3.955  & 2.854  & 6.077  & 5.333\tabularnewline
\hline 
4  & 1.905  & 0.925  & 2.753  & 1.557\tabularnewline
\hline 
5  & 5.062  & 2.884  & 7.734  & 6.247\tabularnewline
\hline 
6  & 3.468  & 1.695  & 5.352  & 3.044\tabularnewline
\hline 
7  & 4.716  & 3.722  & 8.820  & 5.787\tabularnewline
\hline 
8  & 5.087  & 2.647  & 5.625  & 4.817\tabularnewline
\hline 
9  & 6.423  & 4.978  & 7.620  & 7.667\tabularnewline
\hline 
10  & 6.501  & 4.125  & 9.235  & 6.703\tabularnewline
\hline 
11  & 7.446  & 6.703  & 10.767  & 9.400\tabularnewline
\hline 
12  & 6.277  & 6.126  & 11.414  & 8.535 \tabularnewline
\hline 
13  &  & 8.817  & 12.449  & 10.864\tabularnewline
\hline 
14  &  & 8.633  & 12.075  & 10.481\tabularnewline
\hline 
15  &  & 11.558  & 12.285  & 13.636\tabularnewline
\hline 
16  &  & 11.377  & 10.163  & 12.706\tabularnewline
\hline 
17  &  &  &  & 16.113\tabularnewline
\hline 
18  &  &  &  & 15.317 \tabularnewline
\hline 
19  &  &  &  & 18.897\tabularnewline
\hline 
20  &  &  &  & 18.347\tabularnewline
\hline 
21  &  &  &  & 22.136\tabularnewline
\hline 
22  &  &  &  & 21.793 \tabularnewline
\hline 
23  &  &  &  & 25.885\tabularnewline
\hline 
24  &  &  &  & 25.532\tabularnewline
\hline 
\end{tabular}

\newpage{}

\noindent %
\noindent\begin{minipage}[t]{1\columnwidth}%
Table 5: Energy levels (in MeV) of odd A nuclei with a Q.Q interaction.   %
\end{minipage}

\hspace{2cm} %
\begin{tabular}{|c|c|c|c|c|}
\hline 
2J  & $^{43}$Sc ,$^{53}$Fe,$^{53}$ Co  & $^{97}$Cd  & $^{95}$Ag  & $^{93}$Ag\tabularnewline
\hline 
\hline 
1  & 3.906  & 7.159  & 7.623  & 6.833 \tabularnewline
\hline 
3  & 3.284  & 6.585  & 6.164  & 5.365\tabularnewline
\hline 
5  & 2.018  & 5.485  & 4.426  & 4.075\tabularnewline
\hline 
7  & 0.000  & 3.441  & 0.000  & 0.000\tabularnewline
\hline 
9  & 0.816  & 0.000  & 0.997  & 0.740\tabularnewline
\hline 
11  & 1.905  & 1.602  & 2.250  & 1.666\tabularnewline
\hline 
13  & 3.217  & 3.156  & 3.361  & 2.531\tabularnewline
\hline 
15  & 3.467  & 4.752  & 5.154  & 3.708 \tabularnewline
\hline 
17  & 4.088  & 5.703  & 5.874  & 4.597\tabularnewline
\hline 
19  & 2.700  & 6.852  & 8.640  & 4.611\tabularnewline
\hline 
21  &  & 6.585  & 8.252  & 5.992 \tabularnewline
\hline 
23  &  & 6.585  & 5.675  & 7.624\tabularnewline
\hline 
25  &  & 4.374  & 8.032  & 8.260\tabularnewline
\hline 
27  &  &  & 10.173  & 10.775\tabularnewline
\hline 
29  &  &  & 11.295  & 10.499\tabularnewline
\hline 
31  &  &  & 13.139  & 13.495\tabularnewline
\hline 
33  &  &  & 12.907  & 12.710\tabularnewline
\hline 
35  &  &  & 14.475  & 15.045\tabularnewline
\hline 
37  &  &  & 12.840  & 15.318 \tabularnewline
\hline 
39  &  &  &  & 18.253 \tabularnewline
\hline 
41  &  &  &  & 18.091\tabularnewline
\hline 
43  &  &  &  & 21.424\tabularnewline
\hline 
45  &  &  &  & 20.761\tabularnewline
\hline 
\end{tabular}

\newpage{}%
\noindent\begin{minipage}[t]{1\columnwidth}%
Table 6: Energy levels (in MeV)  of Odd-Odd Nuclei with a Q.Q Interaction. %
\end{minipage}
\noindent

 \hspace{3cm} %
\begin{tabular}{|c|c|c|c|c|}
\hline 
J  & $^{44}$Sc ,$^{52}$Mn  & $^{48}$V  & $^{94}$Ag  & $^{96}$Ag\tabularnewline
\hline 
\hline 
0  & 2.982  & 5.623  & 0.000  & 3.506\tabularnewline
\hline 
1  & 0.000  & 0.000  & 0.275  & 0.000\tabularnewline
\hline 
2  & 0.474  & 0.001  & 0.631  & 0.388\tabularnewline
\hline 
3  & 0.960  & 0.558  & 1.147  & 1.037\tabularnewline
\hline 
4  & 1.786  & 0.308  & 1.885  & 1.823\tabularnewline
\hline 
5  & 2.066  & 0.588  & 2.731  & 2.694\tabularnewline
\hline 
6  & 0.472  & 0.914  & 3.667  & 3.511\tabularnewline
\hline 
7  & 1.779  & 1.426  & 0.657  & 3.779\tabularnewline
\hline 
8  & 2.989  & 2.564  & 2.217  & 0.589\tabularnewline
\hline 
9  & 3.427  & 2.682  & 3.810  & 2.580\tabularnewline
\hline 
10  & 5.254  & 4.446  & 5.555  & 4.511\tabularnewline
\hline 
11  & 4.450  & 4.407  & 6.741  & 5.727\tabularnewline
\hline 
12  &  & 6.412  & 9.014  & 7.544\tabularnewline
\hline 
13  &  & 6.520  & 9.189  & 7.409\tabularnewline
\hline 
14  &  & 9.079  & 9.720  & 9.013\tabularnewline
\hline 
15  &  & 9.263  & 11.127  & 7.245\tabularnewline
\hline 
16  &  &  & 13.312  & \tabularnewline
\hline 
17  &  &  & 13.460  & \tabularnewline
\hline 
18  &  &  & 15.805  & \tabularnewline
\hline 
19  &  &  & 15.437  & \tabularnewline
\hline 
20  &  &  & 17.745  & \tabularnewline
\hline 
21  &  &  & 16.507  & \tabularnewline
\hline 
\end{tabular}

\newpage
\noindent %
\noindent\begin{minipage}[t]{1\columnwidth}%
Table 7: Energy levels (in MeV)  of even A Nuclei with f$_{7/2}$$^{42}$Sc
and f$_{7/2}$$^{54}$Co interactions. %
\end{minipage}. 
\begin{tabular}{|c|c|c|c|c|}
\hline 
J  & $^{44}$Ti (with f$_{7/2}$$^{42}$Sc )  & $^{52}$Fe (f$_{7/2}$$^{54}$Co)  & $^{48}$Cr (with f$_{7/2}$$^{42}$Sc )  & $^{48}$Cr (f$_{7/2}$$^{54}$Co) \tabularnewline
\hline 
\hline 
0  & 0.000  & 0.000  & 0.000  & 0.000\tabularnewline
\hline 
1  & 5.660  & 5.459  & 5.472  & 5.172 \tabularnewline
\hline 
2  & 1.159  & 1.024  & 1.203  & 1.084 \tabularnewline
\hline 
3  & 5.783  & 5.810  & 5.746  & 5.614\tabularnewline
\hline 
4  & 2.787  & 2.611  & 2.249  & 1.965\tabularnewline
\hline 
5  & 5.868  & 6.234  & 4.302  & 4.251 \tabularnewline
\hline 
6  & 4.065  & 3.989  & 3.484  & 3.062\tabularnewline
\hline 
7  & 6.040  & 5.880  & 5.954  & 5.535\tabularnewline
\hline 
8  & 6.084  & 5.649  & 5.002  & 4.262 \tabularnewline
\hline 
9  & 7.989  & 7.737  & 6.989  & 6.267 \tabularnewline
\hline 
10  & 7.390  & 6.611  & 6.447  & 5.401\tabularnewline
\hline 
11  & 9.871  & 8.617  & 8.623  & 7.671\tabularnewline
\hline 
12  & 7.708  & 6.413  & 7.891  & 6.606\tabularnewline
\hline 
13  &  &  & 11.578  & 10.168\tabularnewline
\hline 
14  &  &  & 10.263  & 8.580\tabularnewline
\hline 
15  &  &  & 14.550  & 12.432\tabularnewline
\hline 
16  &  &  & 13.583  & 11.421\tabularnewline
\hline 
\end{tabular}

\noindent
\noindent\begin{minipage}[t]{1\columnwidth}%
Table 8: Energy levels (in MeV)  of odd A nuclei with f$_{7/2}$$^{42}$Sc and
f$_{7/2}$$^{54}$Co interactions. %
\end{minipage}
\hspace{+1cm} %

\begin{tabular}{|c|c|c|}
\hline 
2J  & $^{43}$Sc (with f$_{7/2}$$^{42}$Sc )  & $^{53}$Fe(with f$_{7/2}$$^{54}$Co ) \tabularnewline
\hline 
\hline 
1  & 4.319  & 4.870 \tabularnewline
\hline 
3  & 2.885  & 3.528 \tabularnewline
\hline 
5  & 3.451  & 3.849 \tabularnewline
\hline 
7  & 0.000  & 0.000 \tabularnewline
\hline 
9  & 1.676  & 1.524 \tabularnewline
\hline 
11  & 2.332  & 2.201 \tabularnewline
\hline 
13  & 3.503  & 3.337 \tabularnewline
\hline 
15  & 3.514  & 3.204 \tabularnewline
\hline 
17  & 4.300  & 4.052 \tabularnewline
\hline 
19  & 3.648  & 2.817 \tabularnewline
\hline 

\end{tabular}

\newpage{}%
\noindent %
\noindent\begin{minipage}[t]{1\columnwidth}%
Table 9: Energy levels (in MeV)  of Odd-Odd Nuclei with f$_{7/2}$$^{42}$Sc
and f$_{7/2}$$^{54}$Co interactions. %
\end{minipage}. 
\hspace{+1cm}
\begin{tabular}{|c|c|c|c|c|}
\hline 
J  & $^{44}$Sc (with f$_{7/2}$$^{42}$Sc )  & $^{52}$Mn (f$_{7/2}$$^{54}$Co)  & $^{48}$V (with f$_{7/2}$$^{42}$Sc )  & $^{48}$Mn (f$_{7/2}$$^{54}$Co)\tabularnewline
\hline 
\hline 
0  & 3.055  & 2.784  & 5.200  & 5.975 \tabularnewline
\hline 
1  & 0.427  & 0.446  & 0.450  & 0.497\tabularnewline
\hline 
2  & 0.000  & 0.155  & 0.000  & 0.093\tabularnewline
\hline 
3  & 0.762  & 0.797  & 0.924  & 0.903\tabularnewline
\hline 
4  & 0.719  & 0.792  & 0.157  & 0.000\tabularnewline
\hline 
5  & 1.279  & 1.325  & 0.761  & 0.460\tabularnewline
\hline 
6  & 0.381  & 0.000  & 0.626  & 5.894\tabularnewline
\hline 
7  & 1.275  & 0.866  & 1.339  & 0.913\tabularnewline
\hline 
8  & 3.099  & 2.554  & 2.484  & 1.980\tabularnewline
\hline 
9  & 3.392  & 2.724  & 2.836  & 2.077\tabularnewline
\hline 
10  & 4.801  & 4.191  & 4.610  & 3.820\tabularnewline
\hline 
11  & 4.635  & 3.604  & 4.596  & 3.548\tabularnewline
\hline 
12  &  &  & 6.993  & 5.895\tabularnewline
\hline 
13  &  &  & 6.910  & 5.493\tabularnewline
\hline 
14  &  &  & 8.809  & 7.474\tabularnewline
\hline 
15  &  &  & 9.531  & 7.757 \tabularnewline
\hline 
\end{tabular}

\section{Qualitatve discussion of the Q.Q. tables}

Here we will simply report whether the Q.Q interaction is {\color{black} able to predict} a
spin gap, a weak isomerism or no isomerism. These states are shown
in Tables 4 , 5 and 6. Note that the excitation energies are given for
$\chi^{''}=\chi$ b$^{4}$=1.0. We can adjust this parameter to get
the (possible) isomeric state at the right energy. 

In Table 4, we consider the even-even nuclei. For four particles ($^{44}$Ti)
and four holes ($^{52}$Fe) we get identical spectra with any interaction
including Q.Q. We get a prediction of a spin gap since J=12$^{+}$
lies lower than J= 11$^{+}$ or J=10$^{+}$. For the eight particle system
$^{48}$Cr we do not predict any isomerism. We get analogous behavior
in the g$_{9/2}$ shell-- a spin gap for the four hole system ( $^{96}$Cd)
 but none for the eight hole case ($^{92}$Pd) or the eight particle case
($^{88}$ Ru).

In Table 5, we consider even-odd and odd-even nuclei. For the three particle
and three hole systems Q.Q gives a spin gap for the J=19/2$^{-}$ {\color{black} state, which  
lies lower} than J=17/2$^{-}$ or 15/2$^{-}$. This pertains to $^{43}$Sc, $^{43}$Ti
, $^{53}$Fe and $^{53}$Co. Q.Q also yields an spin gap for the three
hole system in the g$_{9/2}$ shell $^{97}$Cd and $^{97}$In. The J=25/2$^{+}$
state lies lower than 23/2$^{+}$ or 21/2$^{+}$. However, no isomerism
is forthcoming to the silver isotopes $^{95}$Ag or $^{93}$Ag.

In Table 6, the odd-odd nuclei are considered and there is no isomerism
with Q.Q for any of them.

\section{Discussion of the tables}

A spin gap in $^{52}$Fe was found and studied by D.A. Geesaman et
al. \cite{Geesaman}. The J= 12$^{+}$ state was below the 10$^{+}$. A key
finding pertaining to isomers in the g$_{9/2}$ shell is contained
in the work of Nara Singh et al \cite{Nara}. They found a J=16$^{+}$
state in $^{96}$Cd which was lower in excitation energy than the
lowest J=15$^{+}$ and J=14$^{+}$ states. Thus the 16$^{+}$ could
not decay by magnetic dipole or electric quadrupole radiation.

A popular but somewhat arbitrary definition of a nuclear isomeric
state is one that lives longer than 1 ns. We adopt this definition
here. In Table 10 we show data on half-lives of isomers gathered from
the NNDC \cite{ENSDF} and  corresponding calculations are shown in Tables 12 and 13.
{\color{black} Direct comparison of the calculated half-life with the experimental data} is difficult because the half-lives are super sensitive to the transition energies.
{\color{black} For comparison}, in Table 10 we show very short half-lives
of non-isomeric states in $^{43}$Sc (J=15/2) and $^{44}$Ti (J= 10).

\noindent %
\noindent\begin{minipage}[t]{1\columnwidth}%
Table 10: Half-lives from the National Nuclear Data Center (NNDC). %
\end{minipage}

\hspace{2cm} %
\begin{tabular}{|c|c|c|c|}
\hline 
Nucleus  & E(keV)  & J  & Half life\tabularnewline
\hline 
\hline 
$^{43}$Sc  & 2988.12  & 15/2$^{-}$  & 5.6 (7) ps\tabularnewline
\hline 
 & 3123.73  & 19/2 $^{-}$  & 472 (4) ns\tabularnewline
\hline 
$^{44}$Ti  & 7671.4  & (10$^{+}$$)$  & 1.87 (35) ps\tabularnewline
\hline 
 & 8039.9  & (12$^{+}$)  & 2.1 (4) ns\tabularnewline
\hline 
$^{52}$Fe  & 6958.0  & 12$^{+}$  & 45.9 (6) s\tabularnewline
\hline 
$^{53}$Fe  & 3040.4  & 19/2$^{-}$  & 2.54 (2) min\tabularnewline
\hline 
$^{94}$Ag  & ?  & (7$^{+}$)  & 0.55 (6) s\tabularnewline
\hline 
 & 6670  & (21$^{+}$)  & 0.40 (4) s\tabularnewline
\hline 
$^{95}$Ag  & 4860.0  & (37/2$^{+}$)  & \textless{} 40 ms\tabularnewline
\hline 
$^{96}$Ag  & ?  & (15$^{+}$, 13$^{-})$  & 0.7 (2) $\mu$s\tabularnewline
\hline 
$^{96}$Cd  & ?  & 16$^{+}$  & 0.29$^{+0.11}_{-0.10}$ s [12] \tabularnewline
\hline 
\end{tabular}

In Table 4 we see clearly that with Q.Q the J=12$^{+}$ state for
two protons and two neutrons and with two proton hole and two neutron holes
is a spin gap isomer. The J=12$^{+}$ lies below J=11$^{+}$ and J=10$^{+}$
(6.277 vs. 7.466 and 6.501). {\color{black} The J=12$^{+}$ become
isomer only when it lies below J=11$^{+}$ and J=10$^{+}$.}
{\color{black} As seen in table 10 the half life of $^{44}$Ti is 1.87ps while that of $^{52}$Fe is 45.9s. Clearly the  J=12$^{+}$ state in $^{44}$Ti is a weak isomer while the J=12$^{+}$ in $^{52}$Fe is a spin gap isomer. The reason for this is that in  $^{52}$Fe the 12$^{+}$ state lies below the lowest 10$^{+}$ state but in $^{44}$Ti J=12$^{+}$ is slightly above J=10$^{+}$. In the single j shell model $^{44}$Ti consists of two protons and two neutrons in the $f_{7/2}$ shell  while $^{52}$Fe consists of two proton holes and two neutron holes. In  the single j shell model the hole-hole interaction  is the same as particle-particle  interaction. Thus if the same interaction is used for the two nuclei then the J=12$^{+}$ states would both  either  be weak isomers or would  both be spin gap isomers. To get around the fact that experimentally one  J=12$^{+}$ state is a weak isomer and the other is a spin gap isomer.}
This is done in Table 7 where for $^{44}$Ti
we use as input the spectrum of $^{42}$Sc while for $^{52}$Fe we
use the spectrum of $^{54}$Co. When this is done, we get a spin gap
isomer for $^{52}$Fe but not for $^{44}$Ti. {\color{black} In the $^{52}$Fe 
case we still get an isomer because the J= 12$^{+}$ (6.413 MeV) state is lies below to J=
10$^{+}$ (6.611 MeV) but the half life is much smaller.}

There is a similar story for $^{43}$Sc and $^{53}$Co ($^{53}$Fe).
In Table 5 Q.Q predicts a spin gap isomer but, as seen in Table 8
the local interactions predict that only for A= 53 will there be a
spin gap. The latter {\color {black} two} are in agreement with experiment. There is
a weaker isomerism in $^{43}$Sc because the 19/2$^{+}$ state is
close to the 15/2$^{+}$.

In the g$_{9/2}$ region the J=16$^{+}$ state in $^{96}$Cd is predicted
to be isomeric with the Q.Q interaction, in agreement with experiment.
There are no other isomerisms predicted for the even-even nuclei in
Table 4. This is in accord with experiment and with calculations with
more realistic interactions in larger shell model spaces.

In Table 10, we gathered experimental data of half-lives corresponding  to all cases from Tables 4 to 9 where either there is a calculated spin gap isomer with Q.Q or an isomer due to
a low energy transition. corresponding calculated half-lives {\color{black} are discussed below.}

In Table 11, we show spin gap isomers as predicted by the Q.Q interaction.
For these there cannot be any E2 or M1 decays. We do not attempt to
calculate their lifetimes.

In Table 12, we show, for the most part, calculations of B(E2)'s and
half-lives for cases where J to (J-2) transitions are allowed but
the states are long lived because the energy differences are small.
We also include $^{94}$Ag, although {\color{black} calculated } lifetime is very long.


There is a previously discovered {\color{black} isomeric J= 21$^{+}$ state
in $^{94}$Ag by I. Mukha et al.\cite{Mukha}}
However, with the Q.Q interaction
as seen in Table 12 although the J=21$^{+}$ is lower than J=20$^{+}$,
it lies above J=19$^{+}$. 
{\color{black}  In the absence of empirical data  we have given the Q.Q results for $\chi^{''}$ = 1. }
 With Q.Q. the CCGI interaction \cite{Coraggio} and that
of Qi et al. \cite{Qi} the values of $\Delta$E are smaller and the
half lives are longer, but they also allow for E2 transition. However
Mukha et al. \cite{Mukha} state that the J=21$^{+}$ state decays by proton
emission. If we had an interaction for which $\Delta$E was negative,
however small, that would solve the problem. With the CCGI interaction \cite{Coraggio}
we are almost there.

{\color{black} An important point in Table 12 is that the B(E2)'s for a given nucleus with various
interactions are very close for $^{43}$Sc and $^{94}$Ag, {\color{black} however},  the half-lives are different because of 
difference in the transition energies.}

\noindent\begin{minipage}[t]{1\columnwidth}%
Table 11: Spin gaps with the Q.Q interaction. %
\end{minipage}

\noindent %
\begin{tabular}{|c|c|c|c|c|c|}
\hline 
 & $^{52}$Fe(12)  & $^{53}$Co (19/2)  & $^{97}$Cd(25/2)  & $^{96}$Cd (16)  & $^{96}$ Ag (15)\tabularnewline
\hline 
\hline 
J  & 6.277  & 2.700  & 4.374  & 10.163  & 7.245\tabularnewline
\hline 
J-1  & 7.466  & 4.088  & 6.585  & 12.585  & 9.013\tabularnewline
\hline 
J-2  & 6.501  & 3.467  & 6.585  & 12.075  & 7.409\tabularnewline
\hline 
\end{tabular}

\newpage{}%
\vspace{10mm}

\noindent\begin{minipage}[t]{1\columnwidth}%
Table 12: Selected lifetime calculations. %
\end{minipage}

\noindent \hspace{-1cm} %
\begin{tabular}{|c|c|c|c|c|c|c|}
\hline 
Nucleus  & Interaction  & J$_{i}$  & J$_{f}$  & $\Delta$E (MeV)  & B(E2) e$^{2}$fm$^{4}$  & $\tau$$_{1/2}$ (SM)\tabularnewline
\hline 
\hline 
$^{43}$Sc  & Q.Q  & 19/2$^{-}$  & 15/2$^{-}$  & 0.768  & 5.918  & 358 ps\tabularnewline
 & $^{42}$Sc/$^{54}$Co  &  &  & 0.134  & 5.919  & 2.21 $\mu$s\tabularnewline
\hline 
$^{44}$Ti  & Q.Q  & 12$^{+}$  & 10$^{+}$  & 0.224  & 21.85  & 45.97 ns\tabularnewline
 & $^{42}$Sc  &  &  & 0.317  & 18.03  & 9.82 ns\tabularnewline
 & $^{54}$Co  &  &  & 0.197  & 22.32  & 85.56 ns\tabularnewline
\hline 
$^{95}$Ag  & Q.Q  & 37/2$^{+}$  & 33/2$^{+}$  & 0.068  & 77.35  & 4.990 $\mu$s\tabularnewline
 & CCGI  &  &  & 0.012  & 70.15  & 0.032 s\tabularnewline
 & Qi  &  &  & 0.099  & 70.90  & 0.83 $\mu$s\tabularnewline
\hline 
$^{94}$Ag  & Q.Q  & 21$^{+}$  & 19$^{+}$  & 1.071 & 68.30  & 6.500 ps \tabularnewline
 & CCGI  &  &  & 0.126  & 68.85  & 0.257 $\mu$s\tabularnewline
 & Qi  &  &  & 0.290  & 68.89  & 3.98 ns\tabularnewline
\hline 
\end{tabular}

\noindent\begin{minipage}[t]{1\columnwidth}%
Table 13: Selected lifetime calculations by taking energy difference from the experimental data.%
\end{minipage}

\noindent \hspace{-1cm} %
\begin{tabular}{|c|c|c|c|c|c|c||c|}
\hline 
Nucleus  & Interaction  & J$_{i}$  & J$_{f}$  & $\Delta$$E_{expt}$ (MeV)  & B(E2) e$^{2}$fm$^{4}$  & $\tau$$_{1/2}$ (SM) & $\tau$$_{1/2}$ (Expt.) \tabularnewline
\hline 
\hline 
$^{43}$Sc  & Q.Q  & 19/2$^{-}$  & 15/2$^{-}$  & 0.136 & 5.918  & 2.041 x $10^{3}$ ns  & 472 (4) ns\tabularnewline
 & $^{42}$Sc/$^{54}$Co  &  &  & 0.136  & 5.919  & 2.040 x $10^{3}$ ns   &   \tabularnewline
\hline 
$^{43}$Ti  & Q.Q  & 19/2$^{-}$  & 15/2$^{-}$  & 0.115 & 23.63  & 1.182 x $10^{3}$ ns   &  556 (6) ns \tabularnewline
\hline 
$^{44}$Ti  & Q.Q  & 12$^{+}$  & 10$^{+}$  & 0.369  & 21.85  & 3.759 ns  & 2.1 (4) ns   \tabularnewline
 & $^{42}$Sc  &  &  & 0.369  & 18.03  & 4.556 ns  &  \tabularnewline
 & $^{54}$Co  &  &  & 0.369  & 22.32  & 3.680 ns  & \tabularnewline
\hline 

\end{tabular}

In Table 13, we used calculated B(E2)'s but experimental transition energies. When compared with the experiment the results for $^{44}$Ti is in reasonable agreement. We have done this only for non-spin gap isomers.

\section{Summary: }

In the present work we have performed shell model calculations using
Q.Q interaction in $f_{7/2}$ and then the g$_{9/2}$ space.
This interaction has one parameter, the overall strength. There
is no additional parameter to determine whether a state is isomeric
or not. We have compared this to other empirical interactions. We
find Q.Q is a good predictor of where isomerism will occur, although,
it does not always distinguish between a true spin gap and an isomer
where E2 or M1 is allowed but the energy is very small. 
To get a weak isomer for $^{43}$Sc and a spin gap for $^{53}$Fe 
we have to use different local interactions in each case-one from
the spectrum of $^{42}$Sc and one from the spectrum of $^{54}$Co.

We have predicted half-lives for non-spin gap nuclei $^{43}$Sc, $^{44}$Ti,
$^{94}$Ag and $^{95}$Ag, and shown agreements for spin gap nuclei.
{\color{black} Our predictions are in good agreement when we combined calculated $B(E2)$
values with the experimental energy differences.}
In the case where no information is known, e.g $^{97}$Cd, we predict
a robust spin gap.

\section*{ACKNOWLEDGEMENTS}

P.C.S. acknowledges a research grant from SERB (India), CRG/2019/000556.

\section*{References}

\end{document}